\documentclass[useAMS,usenatbib,epsfig]{mn2e}
\usepackage{natbib}
\usepackage{amsmath,amsfonts}
\usepackage{epsfig}

\def\reff@jnl#1{{\rm#1\/}}

\def\aj{\reff@jnl{AJ}}                  
\def\araa{\reff@jnl{ARA\&A}}            
\def\apj{\reff@jnl{ApJ}}                        
\def\apjl{\reff@jnl{ApJ}}               
\def\apjs{\reff@jnl{ApJS}}              
\def\ao{\reff@jnl{Appl.Optics}}         
\def\apss{\reff@jnl{Ap\&SS}}            
\def\aap{\reff@jnl{A\&A}}               
\def\aapr{\reff@jnl{A\&A~Rev.}}         
\def\aaps{\reff@jnl{A\&AS}}             
\def\azh{\reff@jnl{AZh}}                        
\def\baas{\reff@jnl{BAAS}}              
\def\jrasc{\reff@jnl{JRASC}}            
\def\memras{\reff@jnl{MmRAS}}           
\def\mnras{\reff@jnl{MNRAS}}            
\def\pra{\reff@jnl{Phys.Rev.A}}         
\def\prb{\reff@jnl{Phys.Rev.B}}         
\def\prc{\reff@jnl{Phys.Rev.C}}         
\def\prd{\reff@jnl{Phys.Rev.D}}         
\def\prl{\reff@jnl{Phys.Rev.Lett}}      
\def\pasp{\reff@jnl{PASP}}              
\def\pasj{\reff@jnl{PASJ}}              
\def\qjras{\reff@jnl{QJRAS}}            
\def\skytel{\reff@jnl{S\&T}}            
\def\solphys{\reff@jnl{Solar~Phys.}}    
\def\sovast{\reff@jnl{Soviet~Ast.}}     
\def\ssr{\reff@jnl{Space~Sci.Rev.}}     
\def\zap{\reff@jnl{ZAp}}                        
\def\nat{\reff@jnl{Nature}}             

\title[A search for naphthalene in diffuse interstellar clouds]{A search for naphthalene in diffuse interstellar clouds}

\author[S. Iglesias-Groth ] {Susana Iglesias-Groth,$^{1,2}$, Jonay I. Gonz\'alez Hern\'andez$^{1,2}$, A. Manchado$^{1,2,3}$$\thanks{E-mail: sigroth@iac.es}$ 
\\ 
$^1$ Instituto de Astrofis\'{\i}ca de Canarias, 38200 La Laguna, Tenerife, Canary Islands, Spain \\
$^2$ Dpt. de Astrof\'\i sica, Universidad de La Laguna, E-38205 La Laguna, Tenerife, Spain \\
$^3$ Consejo Superior de Investigaciones Cient\'\i ficas, Madrid, Spain
}  

\date{Accepted Received In original form}
\pagerange{\pageref{firstpage}--\pageref{lastpage}}
\pubyear{2010}

\begin{document}

\label{firstpage}
\maketitle

\begin{abstract}
We have obtained high resolution optical spectroscopy of 10 reddened O-type stars with UVES 
at VLT to search for  interstellar bands of the naphthalene cation (C$_{10}$H$_{8}$$^+$) in the intervening clouds. No absorption features were detected near the laboratory  strongest band of this cation at 6707 \AA~  except for star HD 125241 (O9 I). Additional bands in the optical spectrum of this star  appear to be consistent with other transitions of this cation. Under the assumption that the bands are caused by naphthalene cations
 we derive a column density   N$_{Np^+}$ = (1.2$\pm$ 0.3) x 10 $^{13}$ cm$^{-2}$  similar to the  column density claimed in the Perseus complex star Cernis 52 (Iglesias-Groth et al. 2008).    The strength ratio of the two prominent diffuse
 interstellar bands at 5780 and 5797 \AA~ suggests the presence of a $\sigma$-type cloud in the line of sight of HD 125241.                 
\end{abstract}


\section{Introduction}

The detection of discrete infrared emission bands near 3.3, 6.2, 7.7, 8.6, 11.3 and 12.7 $\mu$m, 
in dusty environments excited by UV photons led to the suggestion that  polycyclic aromatic hydrocarbons (PAHs) were present 
 in the interstellar medium (L\'eger \& Puget  1984,  Allamandola et al. 1985). These  infrared bands are due to C-C and C-H 
 stretching and bending vibrations in an aromatic hydrocarbon material. Since these bands mostly probe specific chemical bonds
  and not any particular molecular structure,  they cannot proivde   unambiguous identification of single PAHs.  
The naphthalene cation (C$_{10}$H$_8$$^+$) is  the most simple  PAH and one of
the best  characterized spectroscopically  in low-temperature gas phase  at
laboratory  (Pino et al. 1999, Romanini et al.  1999). The laboratory
characterization, crucial for a potential identification in the interstellar
medium,  shows that the  strongest optical  band of the naphthalene cation is
located at  6707.4 \AA~ with  a full width at half maximum (FWHM) of approx. 12
\AA. Progressively weaker bands of similar width have been measured at 6488.9,
6125.2 and 5933.5 \AA  (Biennier et al. 2003). Iglesias-Groth et al.  (2008)  reported the detection 
of weak absorption ( less than 1.5 \% of the continuum)  broad optical bands in the spectrum of the 
star Cernis 52 (A3 V, Cernis 1993) which appear to be consistent with the measured
laboratory bands of  the  naphthalene cation. Gonz\'alez Hern\'andez et al. (2009) show
that the detected  bands are too broad to be originated in the photosphere of this star.

\begin{figure*}
\includegraphics[angle=0,width=11cm,height=11cm]{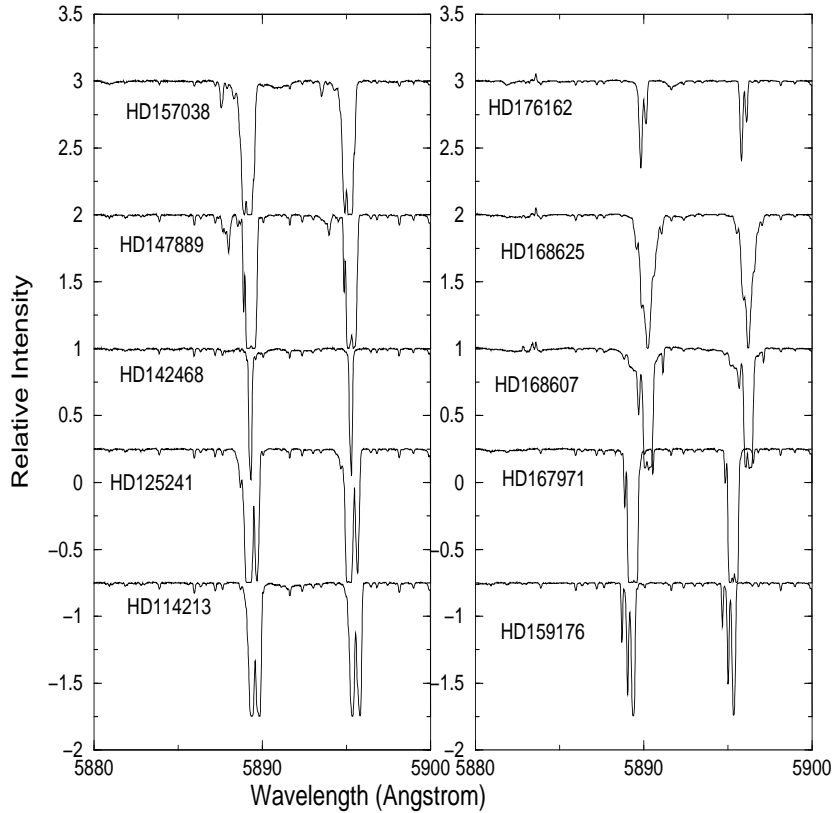}
\caption{ Spectra of the sample stars in the  region of the interstellar Na I doublet }
\end{figure*}

Cernis 52 is located behind or embedded  in a molecular cloud  in the Perseus star forming complex   (one of the nearest to the Sun) that causes  moderate extinction (A$_V$= 3 mag). This cloud presents  millimetric emission (Enoch et al. 2006) and anomalous  microwave (10-70 GHz) emission  (Watson et al. 2005) which  can be caused  by electric dipole radiation of fast spinning PAHs in the intervening cloud (Draine and Lazarian 1998, Planck collaboration 2011m). 
It is possible that this anomalous microwave emission  region in Perseus has an enhanced abundance of PAHs with respect to the diffuse interstellar medium or that the feature is of circumstellar origin and therefore that we are considering a rather peculiar star. Detection of bands of the   naphthalene cation   in other lines of sight  could  show whether  this is indeed an exceptional case.  Alternatively, if this is a common  feature in interstellar clouds,  the hypothesis that some of the  ubiquituous diffuse  interstellar bands (DIBs) are caused by PAHs would be reinforced. So far there is no firm identification of a PAH as responsible of DIBs (Sarre 2006). 

 With the goal to investigate the presence of naphthalene cations in diffuse clouds  we have obtained
  optical spectra of  high resolution and  high signal-to-noise of 10 early-type stars 
with E(B-V) in the range 0.5 to 1.7 (see Table I). There is no report of anomalous microwave emission
 for any of the lines of sight of the stars in our sample.  We detect  the presence of  broad absorption 
 features  in star  HD 125241 with wavelengths and widths  consistent with those of the strongest 
 optical bands of the naphthalene cation. Upper limits to the strength of these bands are set for  
 the other lines of sight.

\begin{table*}
\begin{minipage}{170mm}
\begin{center}
\caption{Basic data of our programme stars.
\label{tab:table_1}}
\scriptsize{
\begin{tabular}{lcccc}
\hline
Star &  Spectral type &E(B-V) & V & S/N\\
HD114213  & B1Ib & 1.11 & 8.97&690\\
HD125241 &09Iab&0.76&8.23&670\\
HD142468 &B0.5IB2& 0.78& 7.88&530\\  
HD147889   &B2III,IV&1.03&7.90&520\\
HD157038 &B1/B2Ia& 0.81&6.71&560\\
HD159176 &O6V&0.36&5.68&224\\
HD166734 &07.5If& 1.36& 8.42&500\\
HD167971  &08,09f&1.04& 7.46&250\\
HD168607&B9Iap&1.65&8.28&400\\
HD176162 &B4V&0.11&5.5 &550 \\
\hline
\end{tabular}
}
\end{center}
\end{minipage}
\end{table*} 

\begin{table*}
\begin{minipage}{170mm}
\begin{center}
\caption{Equivalent widths  (m\AA) of diffuse interstellar bands  for the  star sample.Note:Errors are indicated in parenthesis \
\label{tab:table_2}}
\scriptsize{
\begin{tabular}{lcccccccccccc}
\hline
Star & 5780 & 5797 & 6113  & 6196 & 6203 & 6270 & 6284 & 6376 & 6379 & 6613 &\\

\hline 
HD114213  & 332 (3) & 
            111 (5) & 
	   13 (1)    &    
	  39 (2)&        
	  70 (4)& 
	   42 (5) &           
          540 (60) &      
           31 (2)  & 
	   52 (2)  &  
	   157(5)  &\\

HD125241 &
          526 (10) &
	  109 (5)    &
	   9 (1) &
          56 (2)  &
         131 (5)   &
	  107 (5)&
           1086 (90)& 	
         45 (2)  &
	  69 (2) &
         208 (4) &\\

HD142468 &
               449 (10)  &	  
	     115 (5)  &
	     14 (1)  &
	     55 (2)   &
	     75 (4)       &
	    87 (5) &
	     1176 (90)      &  
	   25 (2)&
	   70 (2) &
	      185 (5)      &\\	
HD147889 &
        347 (8)     &	 
          144 (5)  &  
	 12  (1)  & 
	 39 (2) &
	   80  (5)   &
	  25 (4)&
        389 (50)  &   
	   60 (2)    &	         	  
           86 (2) &
	185 (5)     &\\
	  
HD157038 &
          413  (8)  & 	 	 
 	   93  (5) &
           7   (1) 	&
	   45 (2)&
 	     200 (8)    &
	   74 (5) &
	   1147 (90)   &
	  26 (2) &
	  136 (4) &
	    161 (5)   & \\  
HD159176 &
  161 (5)&
   41 (4)&
   7 (2)&
   25 (2)&
   31 (3)&  
   25 (4)&
   224 (30)&
   7 (1)&
   29 (2)&
   66 (4)& \\
	  
HD166734 &
        687 (15)      &  	
	  250 (8)       &
          48  (3)     &
	  89 (4)&
	  227 (8)         &
	181 (8)    &
	  1209 (90)          &
            99 (8)        	& 
	  220 (2)   &
            398 (7)           &\\
	    
HD167971 &
          530 (20)     &
	     149 (5)           &
	 31 (3)        &
	 16 (2) &
	 100 (8)             &
	  122 (6)   	&  
	      1073 (95)        &
	   45 (3)        &
	 92 (4)      &
            241 (9)    &\\ 
	      			 
HD168607 &
          796 (36)         &  
	   273 (8)          &
	  25 (2)       & 
	  72 (3)&  	
	   149 (5)       &
	   134 (6)&
	   1398 (95)         &
         57 (2)    &
	  152 (4) &
             342 (8)      &\\

HD168625 &
            818 (25) &	 
	     209 (8)  	&
	     37 (3)  &   
	     88 (4)& 
	     146 (5) &
	    185 (7)  &	 
	    993 (80)  &
	    88 (3)      &
	   191 (3) &
	   434 (7)    &\\
	   
\hline
\end{tabular}
}
\end{center}
\end{minipage}
\end{table*} 
                      

\begin{figure*}                 
\includegraphics[angle=0,width=11cm,height=11cm]{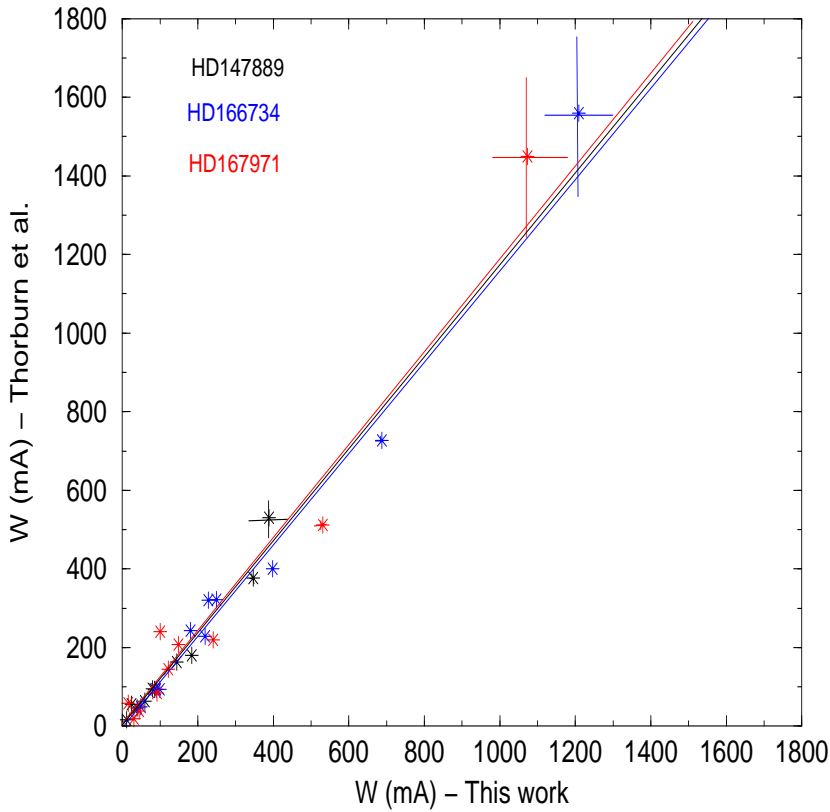}
\caption{ A comparison of DIB equivalent widths  in this work and in  Thorburn et al. (2003).}
\end{figure*}

\begin{figure*}
\includegraphics[angle=0,width=11cm,height=11cm]{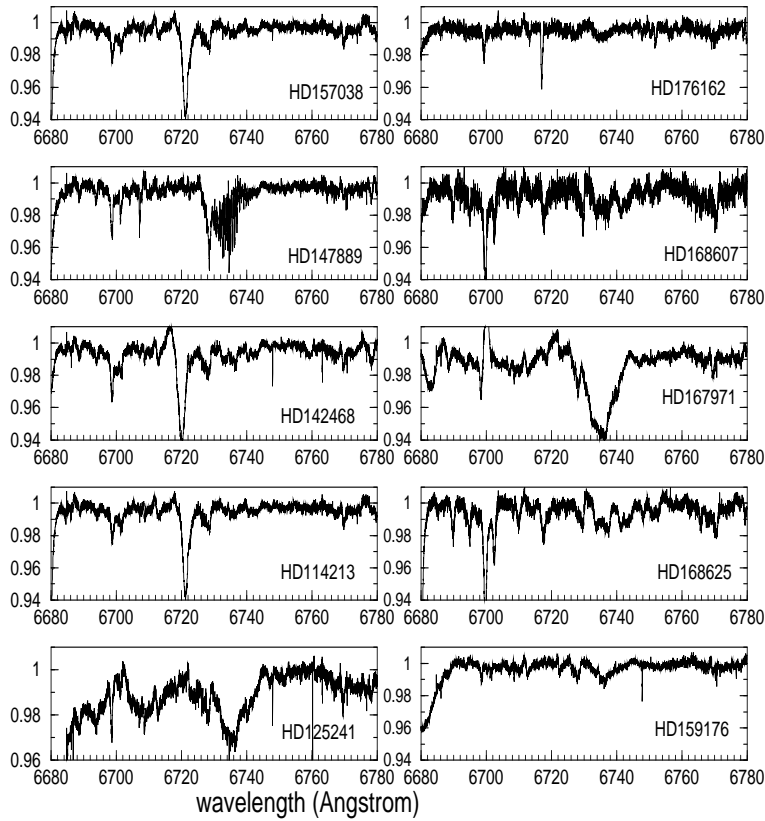}
\caption{ Set of spectra in the range of the most intense band of the naphthalene cation.
 Notice the presence of a band centered around 6708 A in the star HD125241}
\end{figure*}

\begin{figure*}
\includegraphics[angle=0,width=11cm,height=11cm]{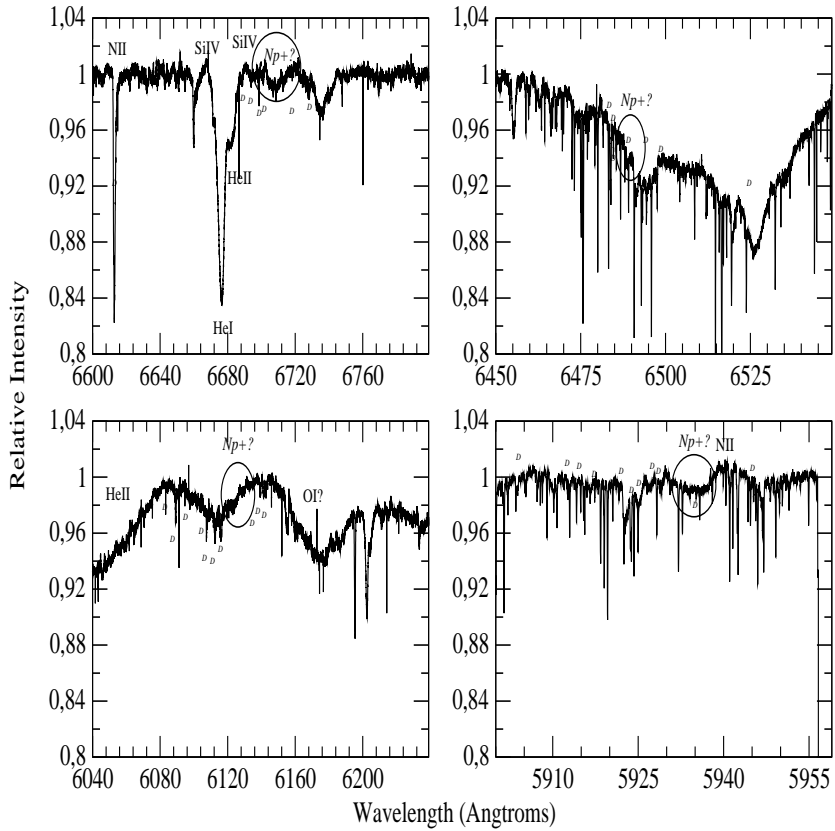}
\caption{Spectrum of HD 125241  in the regions of the four main optical bands of the naphthalene cation (positions indicated by
elipses). Known DIBs in the vicinity of these bands  are marked with "D". Very narrow absorptions are 
due to telluric lines. Broad photospheric
features possibly due to atomic transitions are also indicated.}
\end{figure*}

\begin{figure*}
\includegraphics[angle=0,width=11cm,height=11cm]{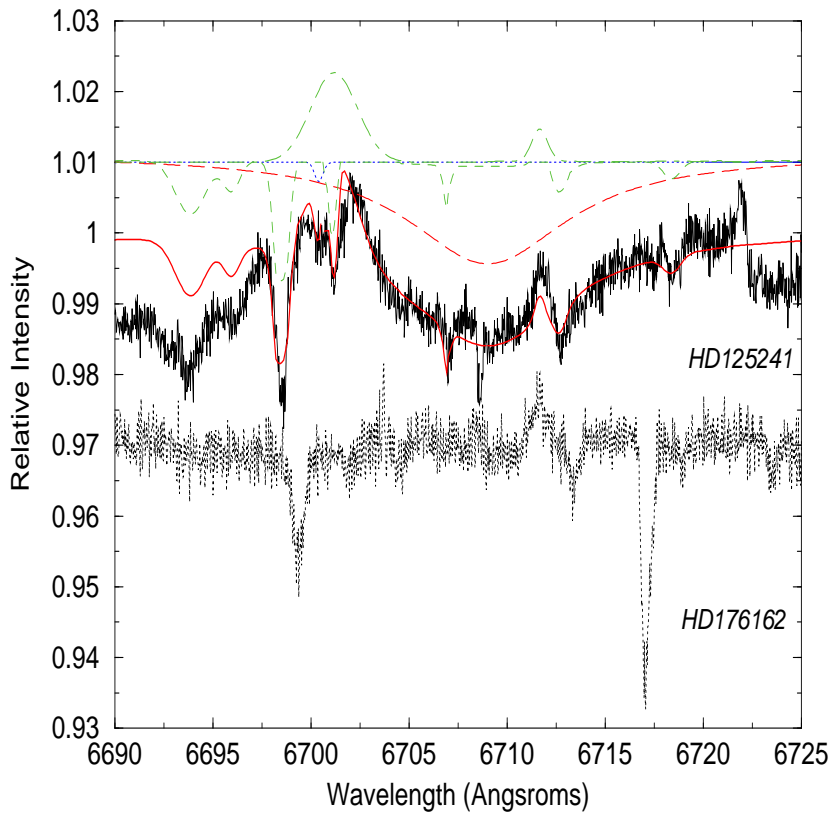}
\caption{Spectrum of HD 125241 in the region of the strongest
band of the naphthalene cation (black solid line).
A model (red solid line) is superimposed to the spectrum.
This model combines a synthetic DIB spectrum (dashed green line),
a synthetic stellar photospheric spectrum (dotted blue line), an amission spectrum probably coming from 
an estended atmosphere (dashed-dotted green line)and a Gaussian of FWHM=12 \AA~ describing
the potential contribution of the naphthalene cation strongest
optical band at 6707.4 \AA~ (red dashed line).
For clarity each of these three contributions are plotted
shifted  by 0.01 in the vertical axis. At the bottom,
the spectrum of star HD 176162 is plotted for comparison.}
\label{fig:f1}
\end{figure*}

\begin{figure}
\center
\includegraphics[scale=0.5]{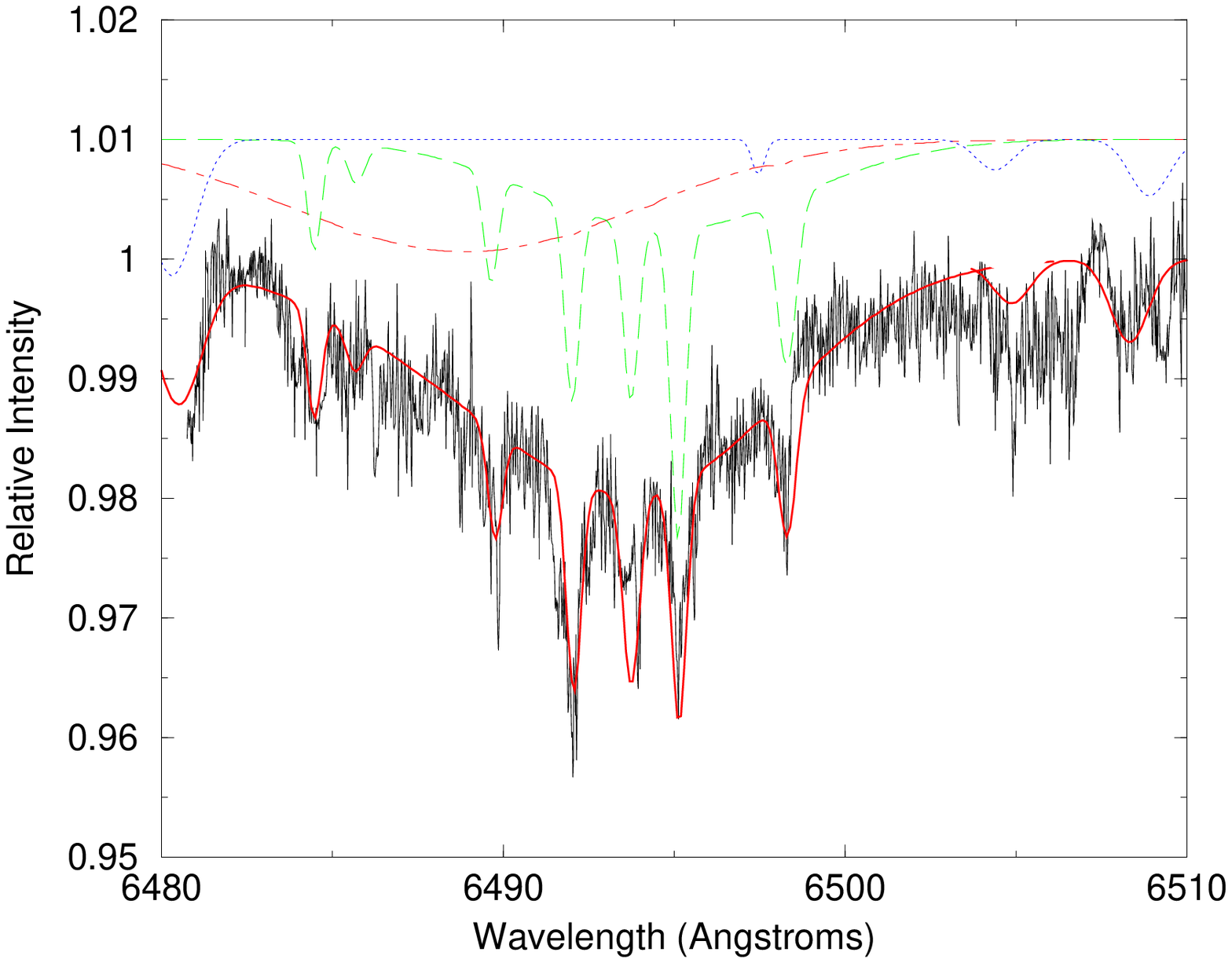} 
\includegraphics[scale=0.5]{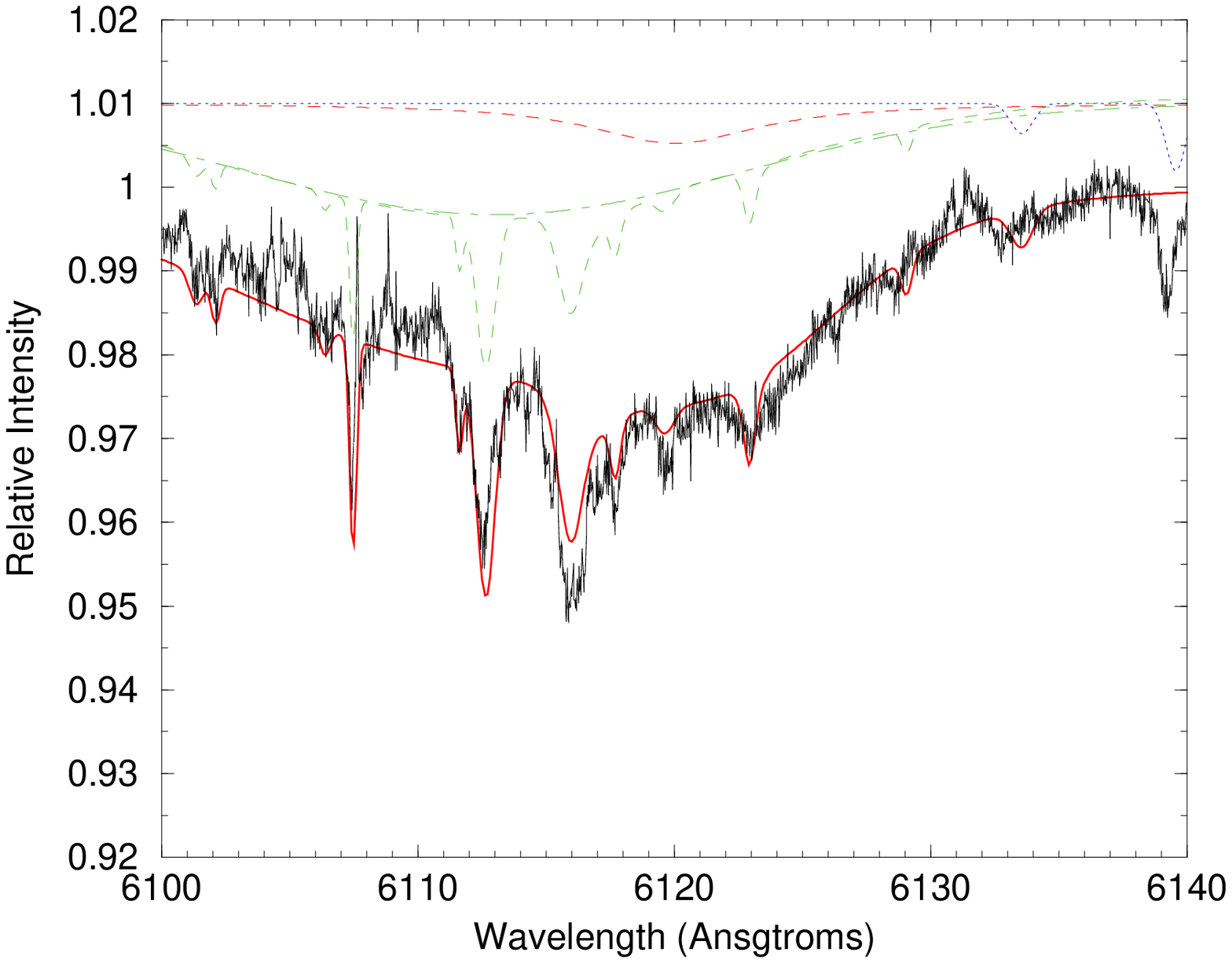} 
\includegraphics[scale=0.5]{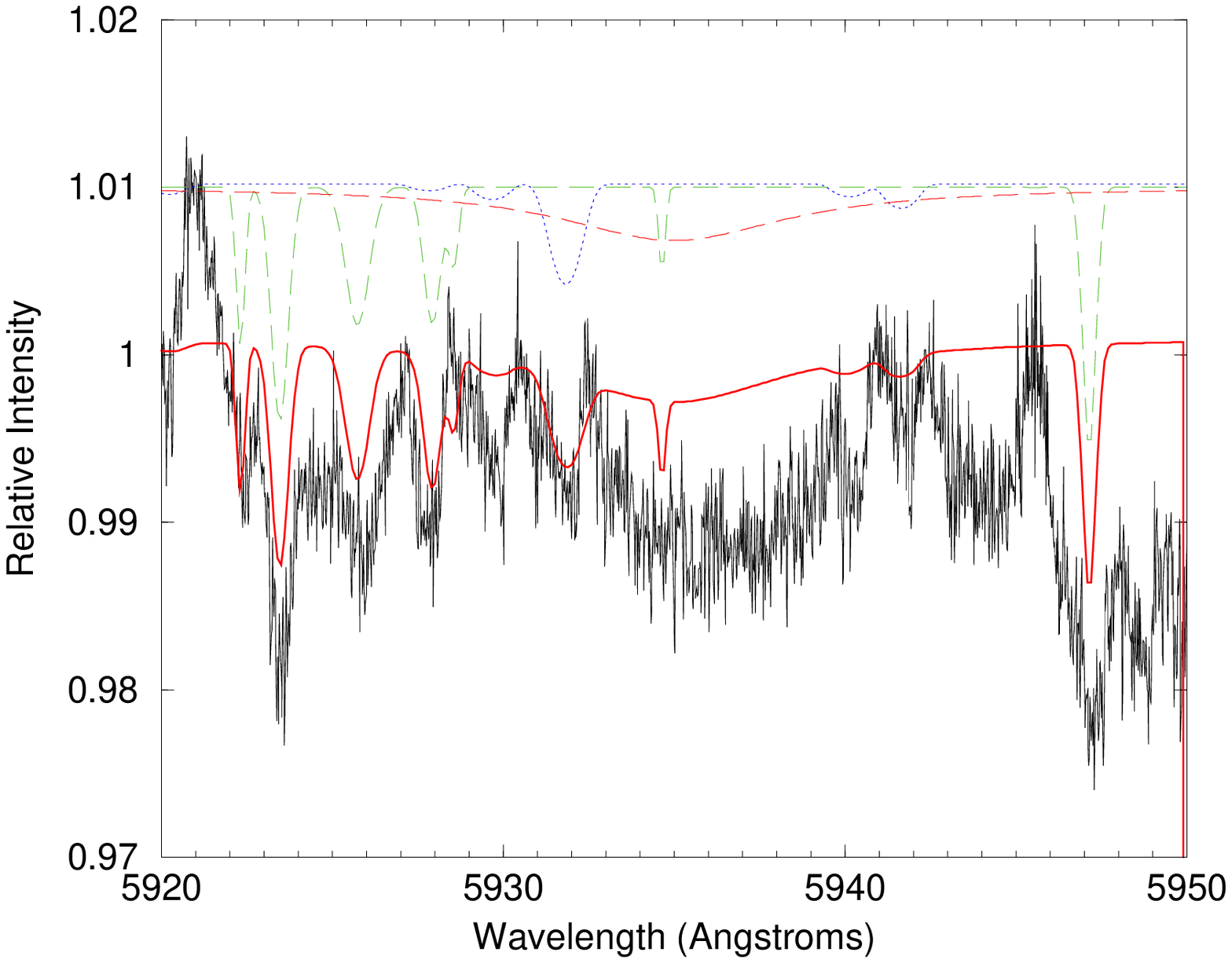}
\caption{\label{fig1}Spectrum of star HD 125241 (solid black line) in the
regions of the 6488.7, 6123.5 and 5934.5 \AA~ naphthalene cation
bands (top, middle and bottom panels). In each panel we plot a
model (red solid line) superimposed to the spectrum. Each model
is the combination of three contributions: DIB synthetic spectrum
(dashed  green line), a stellar photospheric spectrum (dotted blue
line) and a Gaussian of FWHM=10-12 \AA~ describing the potential
contribution of the relevant naphthalene cation band (red dashed line).
In each panel, these individual contributions to the model are
plotted shifted in the vertical axis for clarity.}
\end{figure}

\section{Observations}

The data presented here are based on observations conducted with the UV-Visual Echelle Spectrograph  (UVES) fed by the  VLT (Kueyen unit) 
of the ESO Paranal Observatory, Chile. The observed stars, their V magnitude, their  E(B-V) colour excesses, and the final signal-to-noise 
ratio (S/N) measured near 6750 \AA, are listed in Table 1.  The spectral range  was set to investigate the presence of the four strongest optical  bands of the naphthalene cation.   
Each exposure provided nearly complete spectral coverage from about 5000 to 7000 \AA, at a resolving power 
R$\sim$ 80000.  The maximum S/N per pixel that can be achieved in a single exposure is limited by the nonlinearity of the detector. 
Several spectra (typically between 10 and 20) were obtained for each star in an attempt to achieve the desired S/N in the summed spectrum. 
Total integration times per star ranged from 1 min to 3  min depending on its visual magnitude. 
The many individual exposures   were processed and reduced using  the standard  UVES Data reduction 
package (ESO). 
When a spectral range required correction for telluric lines, each individual spectrum was divided 
by a featureless spectrum of a hot fast rotating star observed with the same instrument configuration.
 The spectra of a given star were combined using the task scombine in IRAF$^¹$ and subsequently 
 normalized to unity. The nomalization was carried out with the task combine of this package using low
  order Legendre polynomials. The continuum regions were selected outside   photospheric features 
  and known diffuse interstellar bands and telluric lines.   Alternative  normalization was also 
  investigated  using a cubic spline fit. The results in both cases were similar.   The profiles of 
  the Na I doublet were examined for the existence of more than one dominating Doppler component
   visible in our high  resolution  and high S/N spectra. Fig. 1 shows this  spectral range for each 
   star in the sample.

\section{Results and Discussion}

In Table 2 we report equivalent widths (W) for a set of well known  DIBs present in 
the spectra of our stars. We used  the IRAF splot task to either  fit a Gaussian
 profile to the band (whenever this was  suitable) or to integrate the band with  
  respect to the pseudocontinuum. The uncertainties of the equivalent widths  were 
  estimated with   the same task, taking into account the S/N in the relevant region 
  of the spectrum and the uncertainty in the location of the continuum  and in the shape
   of the band.  In general these errors (listed in Table 2 are larger than those 
   derived from  the formula $\Delta(W)$= 1.064 x FWHM / (S/N) (see e.g. Hobbs et al.
    2009) which strictly refers to a band of Gaussian shape. We tested our error 
    determination measuring separately in averaged subsets of spectra for each star 
    and calculating  the rms deviation of the equivalent widths.  High resolution spectroscopy for several stars in our sample is available in the literature and 
DIB equivalent widths  have been previously   reported (see e.g. Thorburn et al. 2003).
 A comparison with previous measurements for our stars it  is made in Fig. 2. 
 In general good agreement is found for most of the DIBs with the largest differences in W being 
  of order 20 \%  for the  strongest and broadest  bands where errors are dominated by the uncertainty in 
  the location of the continuum.

In Fig. 3 we plot the final spectra in the region of the strongest optical band of the naphthalene cation. The zero-point of the wavelength scale adopted here is set by assigning the DIB wavelengths 
listed by Hobbs et al. (2008) to the DIBs detected in our spectra. No obvious broad absorptions  are 
detected at the wavelength of the strongest optical band of the naphthalene cation (6707 \AA), except in the 
case of HD 125241 where a broad absorption appears to be present from  approximately  6700 to 6720  \AA. 
HD 125241 is classified as spectral type O9.  We note the similarity of the spectra of  HD 125241 and 
HD 167971 (spectral type O8-9) also plotted in Fig. 3. In the spectral range of this figure, the most
 remarkable features in common between these two stars are: the He II 6683 absorption band and  
 a broad absorption at 6736 \AA~ whose origin is not well established.  There are also  weaker features 
 in both stars  associated with  the DIBs at 6699 and 6729 \AA~ and the SI IV emission line at 
 6701 \AA. HD 167971 has been extensively studied in the  literature, Thorburn et al. (2003) provide 
 equivalent widths for numerous narrow DIBs in this star  which compare well in strength with the 
 values  listed in Table 2.  If the feature at $\sim$ 6707 \AA~ were related to  the most intense band 
 of the naphthalene cation we would expect to find  other bands of this cation  in the spectrum. 
  We  plot in Fig. 4 the spectrum of HD 125241  in a broad region around each of the  four relevant 
   bands. Since the expected widths of the naphthalene bands are of order 12 \AA~ according to  
   laboratory measurements  we plot  spectral ranges at least 10 times larger.  In panel a) we see a 
   complex set of features with  He I in absorption as  the dominant  at 6678 \AA~ and the emission 
   lines of Si IV at 6668 and 6701 \AA~ and the already noticed He II absorption at 6683 \AA.  These 
   Si IV lines are detected in other stars (HD 94963, HD 163758 HD 188001) and are likely formed under 
   non-LTE in an extended atmosphere. We also note  the weak emission line of N II at 6610 \AA~ and 
   that other weak emission lines of N II are present in the 5940-5942 \AA~ range as can be seen in 
   panel d) of the figure.  Outside these emission line regions the spectrum looks normal with many 
   narrow and some broad absorptions that can be mostly abscribed to well known DIBs 
   (listed in Table 2). The other two remarkable features in panel a) are  the absorptions at 6736 
   and 6707 \AA. For the first one, we have no plausible explanation, we just note that in the atlas
    of star HD 183143 by Hobbs et al. (2009) there is a swath of DIBs around this wavelength,
 and for the second, we discuss below if it can be due to the strongest band of the naphthalene cation.

In Fig. 5,  we provide an expanded view of  the spectrum of HD 125241 in the region of the 6707 band and
 compare with  the spectrum of HD 176162, a fast rotating  star with  a rather flat continuum in this 
 region, observed with the same spectrograph configuration.   The comparison with this star, but also
  with other  stars in our sample, indicates that the absorption detected in HD 125241 
  between 6700 and 6720 \AA~ is not due to any systematic effect. The narrow  DIBs 
  at 6699.32, 6702.02 and 6709.43 \AA~ are clearly present in both spectra (note that the 
   wavelength scale has been adopted to match the wavelengths listed by Hobbs et al. (2008) for these 
   DIBs).  In addition we detect the interstellar  lithium doublet at 6707.8 \AA~ in the spectrum of
    HD 125241. This is a  narrow weak feature at the expected wavelength. The total equivalent width 
    of the interstellar Li line is 4$\pm$1 m\AA. The broad absorption between 6700 and 6720 \AA~ seems 
    to be  distorted by  two possible  emission features at approximately  6701 and 6712.5 \AA, 
    respectively,  which appear to be of photospheric origin.  The Si IV  6701 \AA~ emission is 
    significantly stronger with a  bandwidth  of $\sim$ 2 \AA. In the Figure we plot  the synthetic 
    photospheric spectrum of a star with similar spectral type to that of HD 125241 from 
    Tuairisg et al. (2000). The only relevant photospheric feature is an absorption precisely at 6702 \AA, 
    we argue that this feature may appear in emission in our star due to different  physical conditions in 
    its atmosphere. We  reproduce the emission features at 6701 and   6712 \AA~  using  Gaussians. 
    The latter emission line, of uncertain origin,  is also seen in the spectrum of the comparison star 
    plotted at the bottom of Fig. 5.    We also plot a synthetic DIB spectrum using the list  provided by
     Hobbs et al. (2008)  and scaling the features in strength to match those observed in our spectrum.
      The features are scaled individually to obtain a best fit to the observed DIBs  but we note that 
      a similar scaling is required for most of them. All the DIBs in this spectral range are narrow 
      (FWHM$\le$ 0.8 \AA), so  neither the pure photospheric or the pure DIB theoretical spectrum can
       reproduce the observed 6707 broad feature  in HD 125241.  
  We now  explore whether an additional band with the characteristics of the naphthalene cation 6707.4 \AA~ band 
  (FWHM=12 \AA) can provide a better description of the 6700-6720 \AA~ region of the  spectrum. The dashed line in the figure 
	 shows that  the band which would  best fit the observation has a maximum depth 1.4 \%. The continuous 
	 red line is the combination of this naphthalene band plus the stellar 
photospheric spectrum (including the two emission lines mentioned above)  and the DIB synthetic spectrum.
 The combination  provides a good description of the observations  in the region 6700-6720 \AA~ (with 
 reduced $\chi$-squared of 1.06). 
 The total equivalent width of this tentative  cation band would be W(6707)= 240$\pm$50m\AA.  
 We also verified that the adoption of the list of DIBs by Hobbs et al. (2009),which is  based on a 
 different stellar atlas, leads to very similar fits.

If the absorption seen at 6707 \AA~ were due to  naphthalene cations in the interstellar or circumstellar medium in the line of sight of HD 125241, then according to Biennier et al. (2003) we should  expect
the presence of  broad  absorption bands at 6488.9, 6125.2 \AA~ and 5933.5 \AA~  albeit   weaker  than the 
6707 band by a factor 2,4 and 8,  respectively. The precision of the laboratory wavelength  of these bands is 
0.5 \AA.  In Fig.  5 we plot the three relevant spectral regions for HD 125241. Adopting these wavelengths and  a width of 12 \AA~ we
 compute the theoretical naphthalene cation   bands and plot   them in the top, middle and bottom panels of Fig. 5 (dashed lines).   We also account  for  the known DIBs in these  spectral ranges
(green dashed lines) and synthetic photospheric lines (dotted blue lines).
 
 In the range 6480-6500 \AA, there is  a broad absorption in  the spectum of the star. The synthetic photospheric spectrum clearly 
 denotes that most of the absorption in this spectral range is of interstellar or circumstellar origin. A very significant contribution 
comes from the well known DIB at 6494.05 (FWHM$\sim$9 \AA) 
which has been detected in many lines of sight (see e.g. Hobbs et al. 2008, 2009). It is difficult to disentangle the contribution 
of this DIB from the potential absorption at 6489 \AA~produced by the naphthalene cations, but a combination of our predicted 
naphthalene cation band (dashed red line with maximum absorption of 0.7 \%) with the DIB synthetic  model (green continuous line)
 plotted in the figure   seems to reproduce well (with reduced $\chi$-squared of 1.06) the observations in this spectral range.

In order to study the presence of the third strongest band of naphthalene, at the middle panel of Fig. 5 we plot our prediction for this  band scaling from the 6707 \AA~ band (red dashed line) as indicated above. 
After combining with the known DIBs (green dashed line) and the stellar photospheric spectra (dotted blue line)  we compare with the
observations (red solid line).  There are no significant stellar photospheric bands in this spectral range, neither strong DIBs. 
Two narrow DIBs are clearly seen at 6113 and 6117 \AA~ which  are helpful to set precisely the wavelength scale of interstellar 
absoprtions. We produce a synthetic DIB model using  the information in Hobbs et al. (2008) which  reproduces well the two narrow
 DIBs. The observed  spectrum can be reproduced by a combination of these DIBS with two new  broad band features 
  at 6113 and at  6124 \AA~ (with reduced $\chi$-squared of 1.9). The latter  coincides in wavelength and intensity with the values expected for the third band of 
  the naphthalene cation.

Finally, at the bottom panel of Fig. 5. we investigate the presence of the fourth and weakest naphthalene cation band at 5934.5 \AA.  This is a band  particularly difficult  to address since the  
maximum depth  we expect is about 0.2 \% of the continuum.  As in the other panels of the figure  we plot the synthetic photospheric spectrum, the DIB synthetic spectrum and the predicted scaled naphthalene cation band. 
The combination of the three  leads to the continuous red line. The model does provide a poor fit according
to the $\chi$-squared value. A broad absorption band seems to be present in the observed spectrum 
 at the expected location of the fourth naphthalene band, so the presence of naphthalene is possible, 
 but other additional bands of unknown nature are likely contributing to the strength of the
 observed feature.

In summary,  the characteristics of the absorption features  found in the spectral ranges of the 
four strongest optical bands of the naphthalene cation are consistent with the presence of these molecules in the 
intervening cloud  in the line of sight of HD 125241. Adopting  for the oscillator strength of the transition at 6707.4 \AA~ a 
value of f=0.05 (Pino et al. 1999) and 
using the  measured equivalent width of the band, we derive as in  Iglesias-Groth et al. (2008)    a column density of  N$_{Np^+}$ $\simeq$ (1.2$\pm$ 0.3) x 10$^{13}$   cm$^2$. This is a very similar value to that found in the line of sight of Cernis 52. For the remaining stars in our sample we can set upper limits to the column density which vary from one star to other depending on the S/N in the relevant spectral region.
As listed in Table 1 most of the stars have S/N of order 500. This means that bands at 6707 \AA~ with maximum depth 1 \% would have been detected in most cases with high level of confidence. We  set upper limits of 
N$_{Np^+}$ $\le$ 5 x 10$^{12}$   cm$^2$ for the intervening clouds in the other lines of sight.

\begin{figure*}                 
\includegraphics[angle=0,width=11cm,height=11cm]{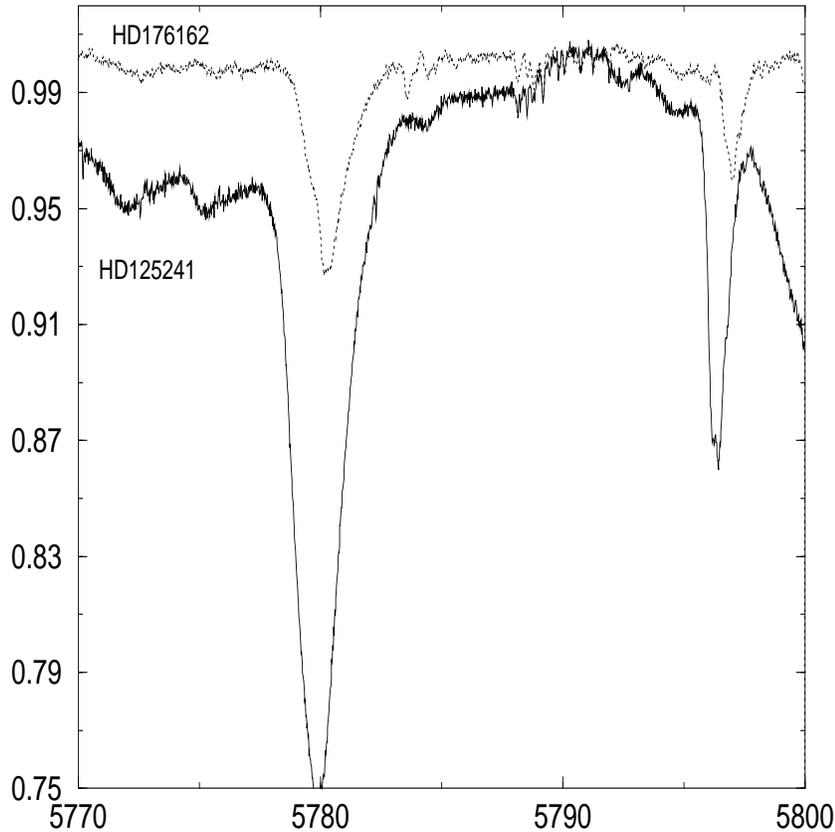}
\caption{ DIBs at 5780 and 5797 \AA~ in the spectra of stars   HD 125241 (solid
line) and HD 176162 (dotted line).We shown relative intensity against of the wavelength (Angstroms) }
\end{figure*}

The extinction in HD 125241 is similar  to that of Cernis 52, but the equivalent widths and profiles 
 of the  5780 and 5797 DIBs in HD 125241 (see Fig. 6)  indicate that the intervening  cloud is 
 of  $\sigma$ type while in Cernis 52 is of $\zeta$-type. It is important to carry out a detailed 
 characterization of the physical parameters and  molecular material in this new cloud.  
 In the case of the intervening cloud towards Cernis 52, spectroscopy of the C$_2$ bands 
 (Iglesias-Groth 2010a) and CH, CH$^+$ (Iglesias-Groth et al. 2010b) indicate a rather high
  abundance of these molecules. The physical and chemical  conditions may be suitable for
  formation of PAHs.  Unfortunately very little is known about the cloud in HD 125241  and 
   CH and CH$^+$ transitions are not covered by our spectra neither  the  strongest series
   of  C$_2$. Only the (4,0) Phillips series is present in our spectrum but the transitions
   are too weak preventing any reliable measurement of the gas  kinetic temperature.   

Interestingly, there are four clouds identified using Spitzer images  in the field of the 
star HD 125241, they are located at angular distances of $\sim$1.3, 5.1, 6.9 and 7.5 arcmin
 from the star  (Peretto \& Fuller 2009). The angular size of the cloud  at a closer angular 
 distance seems to be only $\sim$ 0.1 arcmin, and therefore, without more information it is
  not possible to abscribe the absorption features detected in the spectrum to this cloud. 
  Inspection of IRAS data shows   that HD 125241 is among the stars in our sample with  higher
   dust emission in the light of sight.

Very recent work by Galazutdinov et al. (2011) confirms the presence of a broad band at 6708 \AA~ in 
the spectrum of the star Cernis 52. The significantly  lower S/N of their spectrum with respect the
 one obtained by Iglesias-Groth et al. (2008) at the 9m  HET telescope  prevents that these authors 
 detect  other weaker bands of naphthalene in Cernis 52. They obtained  however  high S/N spectra 
 for other brighter and reddened stars where  they do not find evidence for any broad absorption 
 at 6707-08 \AA. . We do find a 
 similar result in the search carried out here. Only one over 10 stars 
 may present a broad absorption at 6707-08 \AA~ which could be caused by  naphthalene cations.
 Similarly, the search carried out by Searles et al. (2011) has not provided any clear detection of 
 this band in other lines of sight, these authors set stringent upper limits to the column density of
 naphthalene cations and note that a weak broad absorption may exist in just a few cases.
  In summary, the broad feature first detected in Cernis 52 appears 
 to be rather uncommon. While it is possible that in certain  clouds  the column density of these 
 cations is high enough to allow   detection, we cannot discard  the possibility that such feature originates in the circumstellar 
  environment of Cernis 52 and of HD 125241   because of special circumstances that may lead to a
   rich PAH chemistry. It is  important to conduct additional studies of these two stars and their 
   environments in order to establish if this is the case.

\section{Conclusions}

We have searched for  the strongest  optical transition of the naphthalene cation at 6707 \AA~  in the spectra of 
10 reddened O and B type stars.
We find evidence for a broad absorption which could be associated to this band  only in the case of the O9 I star HD 125241 which displays  
a broad absorption band at 6707 \AA~  consistent   in wavelength and FWHM with laboratory measurements of the strongest optical band of the  naphthalene cation.
Weaker bands of this  cation seem to  be present in the spectrum of this star with consistent strength at 6489, 6125 and 5934 \AA. 
Assuming that these bands are indeed caused by the naphthalene cations we derive a column density of  N$_{Np^+}$ = (1.2$\pm$ 0.3) x 10$^{13}$   cm$^2$.
The    diffuse interstellar  bands at 5780 and 5797 support the presence of  a sigma-type cloud in the line of sight of this star.

In the other stars of the sample we set upper limits to this column density a factor 2-3
lower in spite of the higher extinction associated with several of the intervening clouds. 
Additional studies of the  
intervening cloud and circumstellar environment in HD 125241 may provide valuable
information  on the physical and chemical conditions that may govern an active  formation
of PAHs.  

\end{document}